\documentclass[a4paper,fleqn,usenatbib,useAMS]{mnras}

\usepackage[T1]{fontenc}
\usepackage{ae,aecompl,caption,subcaption,xfrac}

\usepackage{graphicx}	
\usepackage{amsmath}	
\usepackage{amssymb}	
\usepackage{float}

\newcommand{\myref}{/Users/Vasiliy/Desktop/PhD/ref}

\def \mpcoh{\,h^{-1}{\rm Mpc}}

\title[Spherical Evolution of Voids]{Testing the spherical evolution of cosmic voids}

\author[Demchenko, Cai et al.]{
Vasiliy Demchenko\thanks{E-mail: vgd@roe.ac.uk},
Yan-Chuan Cai,
Catherine Heymans,
and John A. Peacock
\\
Institute for Astronomy, University of Edinburgh, Royal Observatory, Edinburgh EH9 3HJ, UK\\
}

\date{Accepted 2016 August 10. Received 2016 July 18; in original form 2016 May 17}

\pubyear{2016}

\begin{document}
\label{firstpage}
\pagerange{\pageref{firstpage}--\pageref{lastpage}}
\maketitle

\begin{abstract}
We study the spherical evolution model for voids in $\Lambda$CDM, where the evolution of voids is governed by dark energy at an earlier time than that for the whole universe or in overdensities.  We show that the presence of dark energy suppresses the growth of peculiar velocities, causing void shell-crossing to occur at progressively later epochs as $\Omega_{\Lambda}$ increases. We apply the spherical model to evolve the initial conditions of N-body simulated voids and compare the resulting final void profiles. We find that the model is successful in tracking the evolution of voids with radii greater than $30\mpcoh$,  implying that void profiles could be used to constrain dark energy. We find that the initial peculiar velocities of voids play a significant role in shaping their evolution. Excluding the peculiar velocity in the evolution model delays the time of shell crossing. 

\end{abstract}




\section{Introduction}
The cosmic web, consisting of haloes, voids, filaments, and walls in large-scale structure is predicted by the cold dark matter model \citep{bondetal1996,cosmicweb} and confirmed by large galaxy surveys  \citep[e.g.][]{cfaoriginal,2df,alametal2015}. Among these large-scale structures, the underdensities of the universe, i.e. cosmic voids, have been shown to have great potential for constraining dark energy and testing theories of gravity via several measurements.  These measurements include: distance measurement via the Alcock-Paczy\'nski Test (AP) \citep{ryden1995,lavauxwandelt2012,sutteretal2014}, weak gravitational lensing of voids \citep{krauseetal2013,melchioretal2014,clampittjain2014,gruenetal2015,sanchezetal2016}, the signal of the Integrated Sachs-Wolfe (ISW) effect associated with voids \citep{Sachs1967, granettetal2008,nadathuretal2012,flenderetal2013,Planck2014, Ilic2014, caietal2014, Kovacs2015, PlanckISW2015, aiolaetal2015}, void ellipticity as a probe for the dark energy equation of state \citep{leepark2009,lavauxwandelt2010,bosetal2012,sutteretal2015,pisanietal2015}, void abundances and profiles for testing theories of gravity and cosmology \citep{lietal2012, clampittetal2013, Lam2015, caietal2015, zivicketal2015, Barreira2015, Massara2015}, coupled dark energy \citep{Pollina2016}, the nature of dark matter \citep{Yang2015}, baryon acoustic oscillations in void clustering \citep{Kitaura2015, Liang2015}, and redshift-space distortions in voids \citep{Hamaus2015, Hamaus2016, Cai2016}. Despite their popularity and great potential as a cosmological tool, a gap of knowledge between the evolution of individual voids through simulations and observations versus theory still persists. How voids evolve from the initial conditions and how dark energy or alternative theories of gravity shape this process still lacks a complete analytical understanding. As with the formation history of haloes, the initial conditions and evolution history of voids sets the base for their two fundamental properties: profile and abundance. As these are crucial for constraining cosmological parameters, it is therefore important to bridge the gap between theory and observations. This is the main goal of our study. 
 
The spherical evolution model has commonly been applied in theoretical studies of voids \citep{peebleslss,blumenthaletal1992,sheth}. However, voids are usually assumed to start evolving from a spherical top-hat underdensity or some smooth functional form, which may not be precise descriptions for the initial underdenisties arising from random Gaussian fluctuations.  Also, the analytical solution for the model is only found for the Einstein de-Sitter (EdS) universe. Solutions for the specific regimes of shell-crossing and turn-around in overdensities in a $\Lambda$CDM universe were given in \cite{ekecolefrenk1996}. The condition for shell-crossing in voids is different from that in overdensities making the solution in \cite{ekecolefrenk1996} inapplicable for shell-crossing in voids. All these factors limit the application of the spherical evolution model and make it an unlikely candidate to describe observations or even simulations. In this study, we take steps to extend the model by generalising it to cosmologies with dark energy and by going beyond simple assumptions for the void profile.  Using the evolution equation to evolve initial void profiles from N-body simulations we find that, given the correct initial density and velocity profiles, the spherical model can reproduce late time void profiles from N-body simulations for void radii $> 30\mpcoh$.

During the preparation of our manuscript, \citet{Wojtak2016} posted a paper on a similar topic, studying void properties (e.g. ellipticity, size and density profile) using simulations.  However, our focus in this paper is on comparing void profiles in simulations with the spherical model, so the two studies are complementary.

The structure of this paper is as follows: Section \ref{sec:eds} introduces the spherical model for an EdS cosmology, Section \ref{sec:LCDM} extends the model to $\Lambda$CDM and provides a comparison between the different cosmologies, Section \ref{sec:nbody} compares the theoretical $\Lambda$CDM model to results from N-body simulations and includes a discussion on the impact of peculiar velocities on void profiles, and Section \ref{sec:conclusion} summarizes the study.

\section{The Spherical Model} \label{sec:eds}
\label{sec:spherical} 
The spherical evolution model was originally introduced to model the evolution of overdensities \citep{gunngott1972}.  This model assumes a spherical underdensity $\rho_i$ embedded in an expanding, homogeneous background with density $\bar\rho$. The evolution of each radius is determined by the total mass $M$ contained within the proper radius $R$ via the acceleration equation in the Newtonian regime.\footnote{The Newtonian regime implies that $\dot{R} \ll c$ and $R \ll R_c \sim c/H$.}
The model makes no assumption about the background cosmology with the evolution given as:
\begin{equation}
\frac{\ddot{R}}{R} = \frac{-4\pi G}{3} \sum_{n}(\rho_n + 3p_n),
\label{equation:accnok}
\end{equation}
where $R$ is the proper radius, the double dot indicates the second derivative with respect to proper time $t$, $G$ is the gravitational constant, and $\rho_n$ and $p_n$ are the density and pressure components, respectively, of any contributing component \emph{i.e.}, radiation, matter, dark energy \citep{cosmoprob}.  The same equation, known as the Friedmann equation, applies to an unperturbed region, which yields the expansion history of the universe.  The spherical model has been applied to solve the evolution of overdensities and underdensities \citep[e.g.][]{gunngott1972,peebleslss, liljelahav1991,sheth}.  Using the spherical model, the evolution equation in a $\Lambda=0$ universe becomes:
\begin{equation}
\ddot{R} = -\frac{GM}{R^2}.
\label{equation:acc}
\end{equation}
To solve the above equation, the initial density and velocity profiles are needed. For the case of an overdensity, which eventually collapses and virialises as a halo, the initial density profile is usually taken to be a spherical top-hat and the initial velocity is assumed to be the Hubble flow at the initial time $t_i$.  We will use the subscript $i$ to indicate quantities at the initial time throughout the paper. With these assumptions, the equation can be solved analytically and the solution for the size of the radius as a function of time takes the following parametric form \citep{gunngott1972,liljelahav1991}:
\begin{eqnarray}
  R &=& A(1-\cos\theta), \nonumber\\
  t+T &=& B(\theta - \sin\theta), \nonumber\\
  A^3 &=& GMB^2, 
\label{equation:parsolhalo}
\end{eqnarray}
where $A$, $B$, and $T$ are constants that can be fixed once the initial conditions are fixed and $\theta$ is an indicator of time.  For voids with the same initial settings, the analytical solutions can also be found by taking an inverse top-hat model for the density profile \citep{gunngott1972,liljelahav1991,peebleslss,sheth}:
\begin{eqnarray}
  R &=& A(\cosh\theta -1), \nonumber\\
  t+T &=& B(\sinh\theta-\theta ), \nonumber\\
  A^3 &=& GMB^2.
\label{equation:parsolvoid}
\end{eqnarray}
Note that the parametric solutions above apply to any $\Lambda=0$ universe.  For this study we use the flat EdS cosmology.

\begin{figure}
	\centering
	\begin{subfigure}{0.5\textwidth}
		\includegraphics[width=1.0\linewidth]{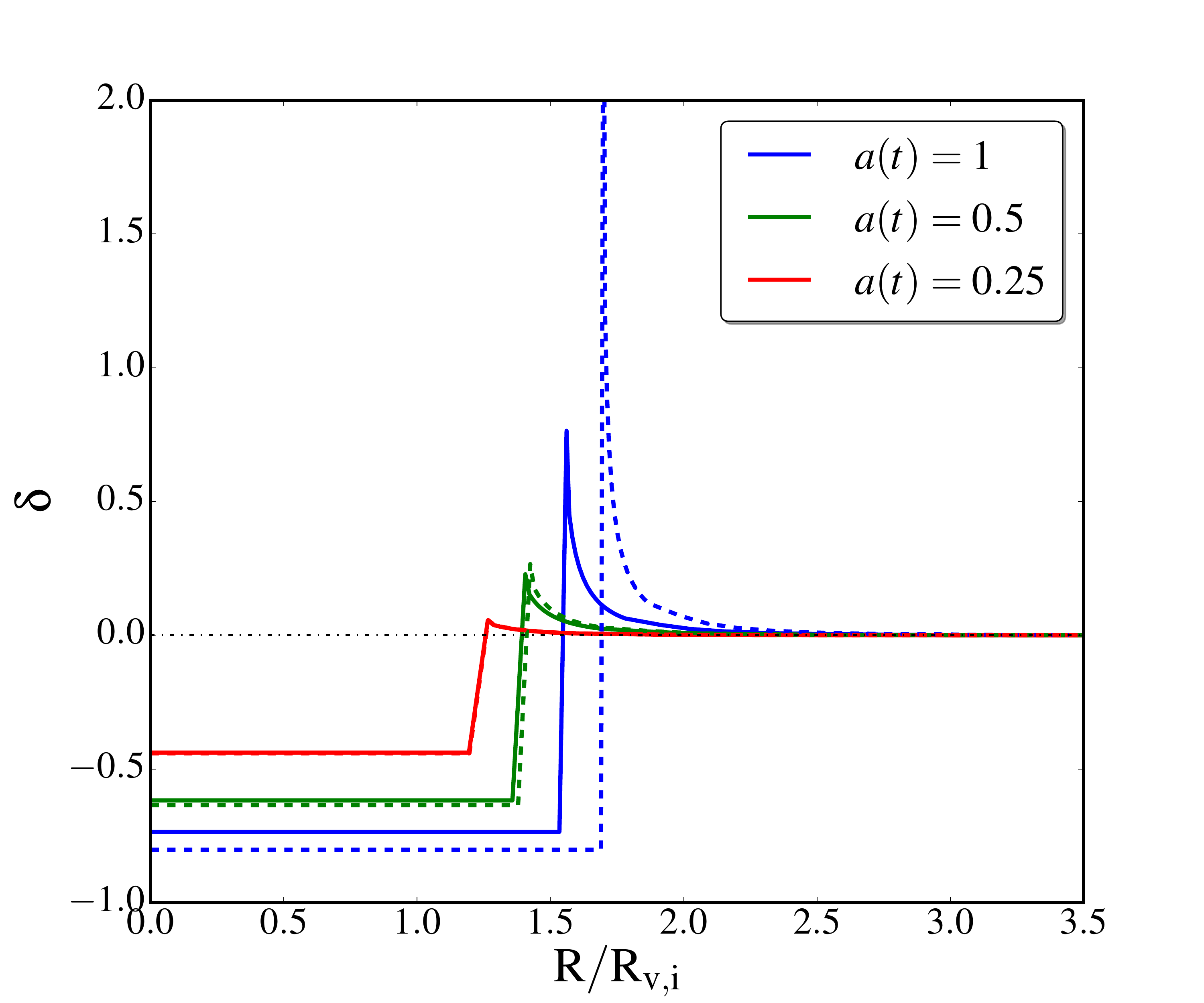}
	\end{subfigure} %
	\begin{subfigure}{0.5\textwidth}
		\includegraphics[width=1.0\linewidth]{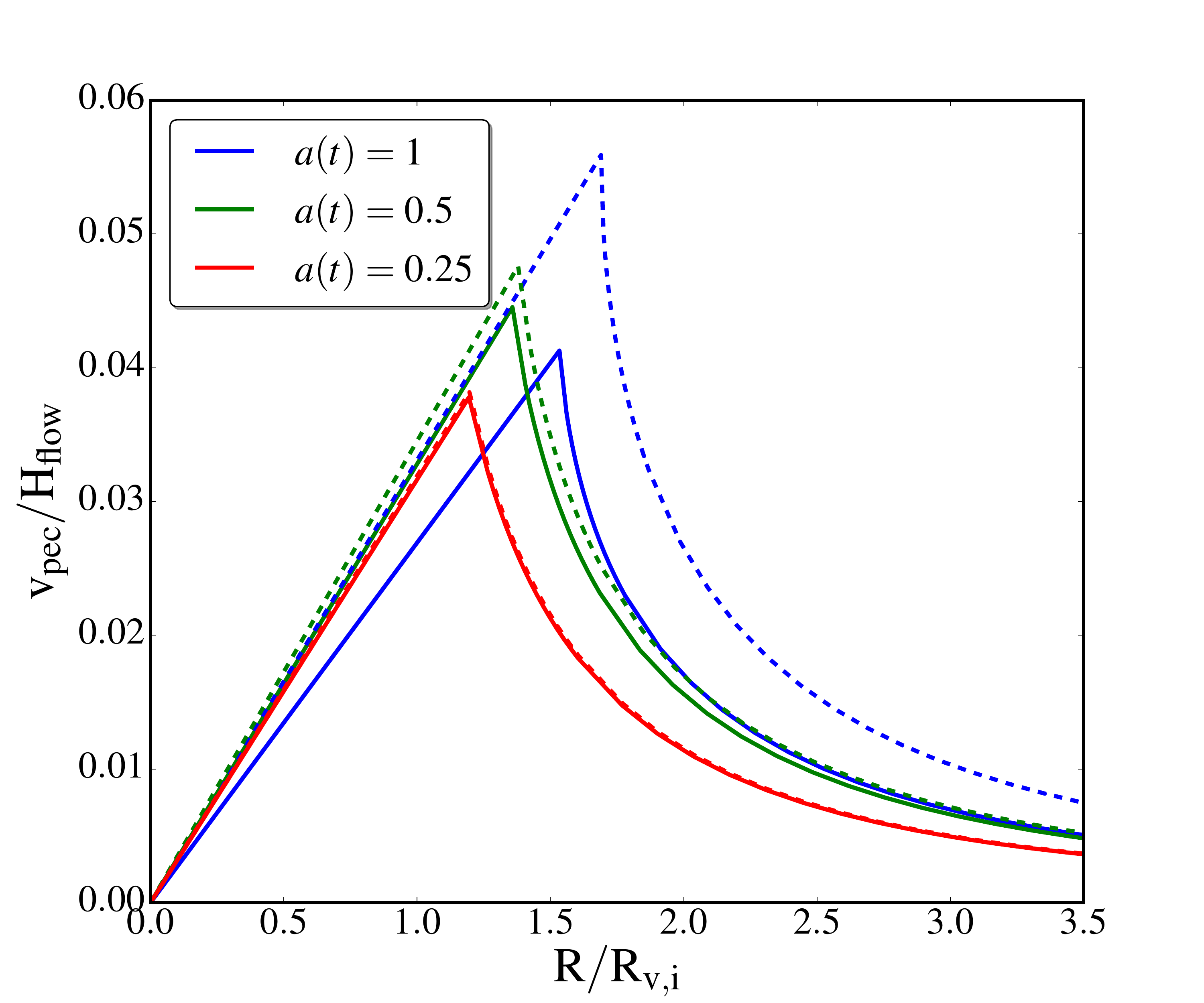}
	\end{subfigure} %
	\caption{Numerically evolved density (top) and proper peculiar velocity (bottom) profiles of a spherical underdensity for both a $\Lambda$CDM (solid) and EdS (dashed) universe. These profiles are shown at three different epochs ($a = 0.25,$ $a = 0.5,$ and $a=1.0$ from left to right) as a function of comoving radius normalised by their initial sizes. Peculiar velocities are normalised by the initial Hubble flow.  The initial density contrast is chosen such that the inner most shell approaches, but not reach, shell crossing for the EdS cosmology.}
	\label{fig:LCDMtot}
\end{figure}

There are noteworthy differences between haloes and voids in this model. For haloes, the overdensity begins expanding with a slower rate than that of the background universe. Since the local density is higher than the background, the effective Hubble rate is higher. The overdensity keeps expanding until it reaches a maximum radius, at which point it turns around and collapses into a singularity. The well-known turnaround radius ($R_{\rm ta} = R_{\rm i}/1.771$) and the density contrast when the over density collapses (linearly extrapolated $\delta_{\rm sc}=1.686$) are found based on these exact assumptions, where $\delta$ is defined as $\delta = \rho/\bar{\rho}-1$.  Note that provided shell-crossing does not occur before turnaround, which is unlikely, these values do not depend on the interior initial density profile. 

\begin{figure}
	\centering
	\includegraphics[width=1.0\linewidth]{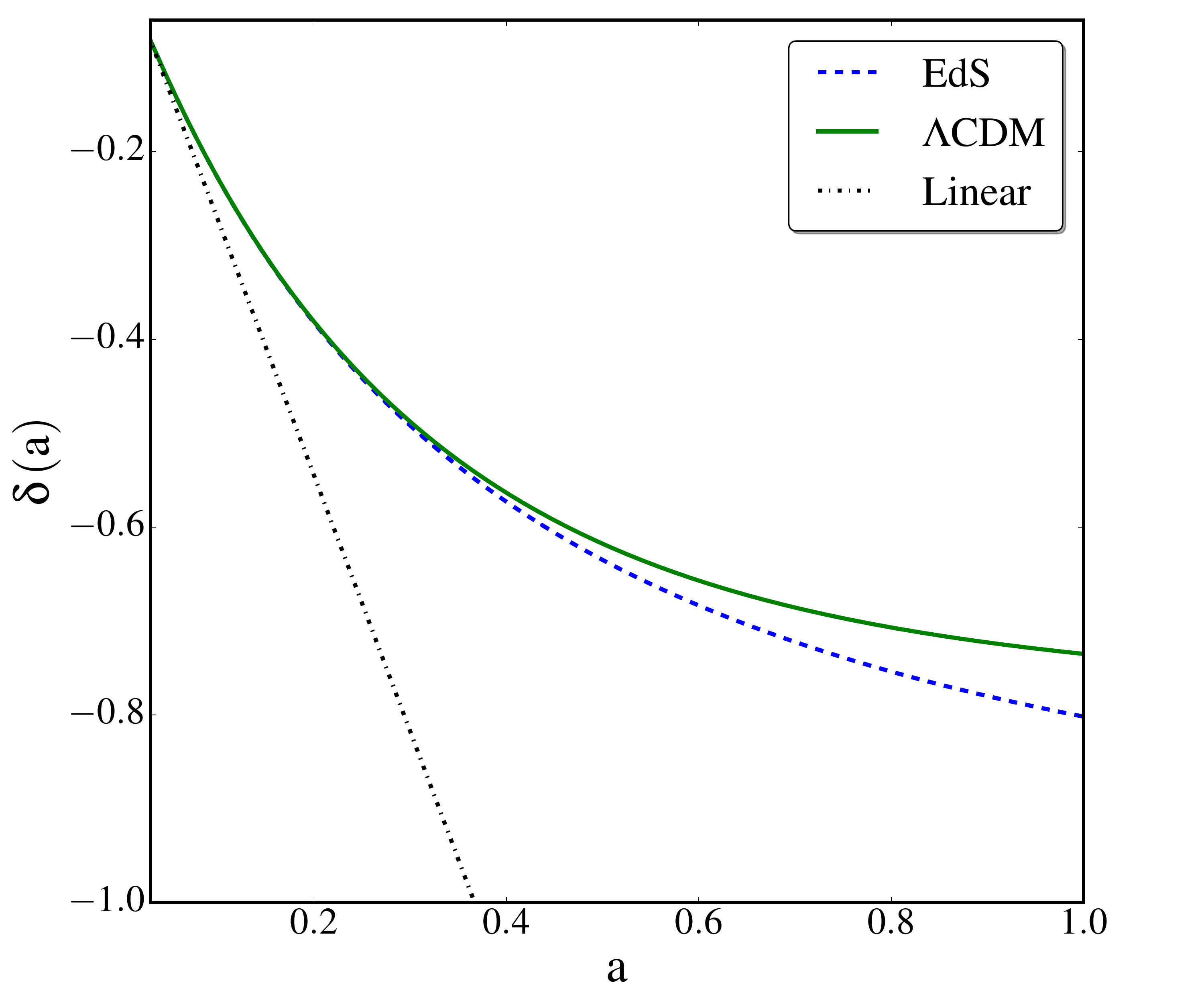}
	\caption{Density contrast of a void versus scale factor for EdS (dashed) and $\Lambda$CDM (solid).  The prediction from linear theory is shown in the dot-dashed line.}
	\label{fig:deltavsa}
\end{figure}

For voids, matter shells will keep expanding from the initial conditions at a faster rate than the background universe. This expansion rate increases as the local density decreases. With this (unrealistic) assumption, a void's expansion is unaffected by its surrounding environment. The expansion of matter shells at radii smaller than the edge of the top-hat, $R_t$, are slightly faster than for those at $R>R_t$. This causes an overdense ridge to build up at the edge of the void. At some point the inner shells catch up with the outer ones. This defines shell-crossing for voids, beyond which the analytical model fails. The evolution of such a case in terms of density contrasts and peculiar velocities is shown in Fig~\ref{fig:LCDMtot}.  In the EdS universe (dashed lines), the comoving radius of the underdensity would have expanded by a factor of 1.7 when shell crossing occurs, and the corresponding density contrast is $\delta=-0.8$, as shown by the dashed curve in Fig~\ref{fig:deltavsa} [See also \citep{blumenthaletal1992,sheth}].  These analytical values are successfully reproduced by our numerical solver for the acceleration equation (Eq \ref{equation:acc}).  

Technically, we achieve the examples given in Figs~\ref{fig:LCDMtot} \& \ref{fig:deltavsa} by numerically solving Eq \ref{equation:accnok} for spherical underdensities, requiring both the initial densities and velocities. We set up an inverse top-hat density contrast $\delta_i$ with a comoving radius of $R_{v,i}$ at the epoch $a_i$, so the mass at a given radius $R$ from the centre is: 
	\[
	M_i(<R_i) = 
	\begin{cases} 
	\frac{4\pi}{3} \bar{\rho_i} R_i^3(1+\delta_i) & \text{if } R_i \le R_{v,i} \\
	\frac{4\pi}{3} \bar{\rho_i} R_i^3\left(1+ \frac{R_{v,i}^3}{ R_i^3}\delta_i \right)   & \text{if }  R_i>R_{v,i}, 
	\end{cases}
	\]
	where $\bar{\rho}$ is the background matter density of the universe.
	For the example of EdS, we integrate Eq \ref{equation:acc} over $t$ once to obtain: 
	\begin{equation}
	\frac{1}{2}\dot{R}^2 = \frac{GM}{R^2}+E,
	\label{equation:vel}
	\end{equation}
	where $M$ is a function of $R$ and $R$ is a function of $t$. The constant of integration $E$ at the initial time $t_i$ is set by the initial kinematic energy, i.e. $E_i=\frac{1}{2}v_i^2$, 
	and the initial total velocities $v_i$  are set to be the same as the Hubble flow, i.e. $v_i=H_iR_i$. We discuss the impact of other choices of initial velocities in Section \ref{sec:nbodycomp}.  Note that in cosmologies with $\Lambda$, there will be a contribution from $\Omega_{\Lambda}$ in the above equation. In non-flat universes, the curvature contributes to the energy term in Eq \ref{equation:vel}, but Eq \ref{equation:accnok} remains the same.

With the above setup, we integrate Eq \ref{equation:vel} for $R(t)$ and use it to solve for the average density contrast within $R$, $\Delta(a,<R)$, defined as: 
\begin{equation}
	1+\Delta(a,<R) = M(<R) \big/ \frac{4 \pi}{3} R^3 \bar{\rho},
	\label{equation:cumdelta}
\end{equation}
and we see that $1+\Delta(a,<R) \propto \left(a/R \right)^3$.  We then differentiate $1+\Delta(a,<R)$ to obtain the density contrast of each spherical shell at $R$, $\delta(a,R)$.
We track  the evolution of 30 consecutive shells equally spaced from the void centre to $3.5 \times R_{v,i}$ from $a_{\rm i} = 0.01$ to $a=1$. We choose an initial density contrast $\delta_i$ such that the void approaches, but does not enter, the shell-crossing regime in an EdS universe.  We solve the background expansion history $a(t)$ with the same setup, apart from setting $\delta_i$=0 and an arbitrary choice of radius. We note that for all the theoretical calculations in the different cosmologies used in this paper, their initial conditions are equal at a fixed $a_i$ and follow the same framework to solve the acceleration equation.  We find excellent agreement between the EdS results from our numerical solver and the analytical EdS solution, providing a benchmark from which we generate void models for other cosmologies.

The density contrast at shell-crossing, $\delta=-0.8$, has been taken as a default choice of theoretical density threshold \citep{sheth}. It is worth noting that the acceleration equation and the form of the solutions are general to any initial density profiles for both voids and haloes, i.e. the top-hat profile assumption need not to be taken. However, quantitatively, the shell-crossing time and density contrast $\delta=-0.8$ are relevant when assuming an inverse spherical top-hat density profile and an EdS universe. Relaxing any of those assumptions may lead to changes in those values. The sharp transition at the edge of the top-hat is somewhat unnatural and unrealistic. The time and density contrast for shell-crossing is likely to be altered if a different (slower varying) initial density profile is assumed. It is the main goal of our paper to test the performance of the spherical evolution model by going beyond these overly simplistic assumptions.   

\section{Spherical Model Extended to $\Lambda$CDM and Beyond} \label{sec:LCDM}
\begin{figure}
	\centering
	\includegraphics[width=1.0\linewidth]{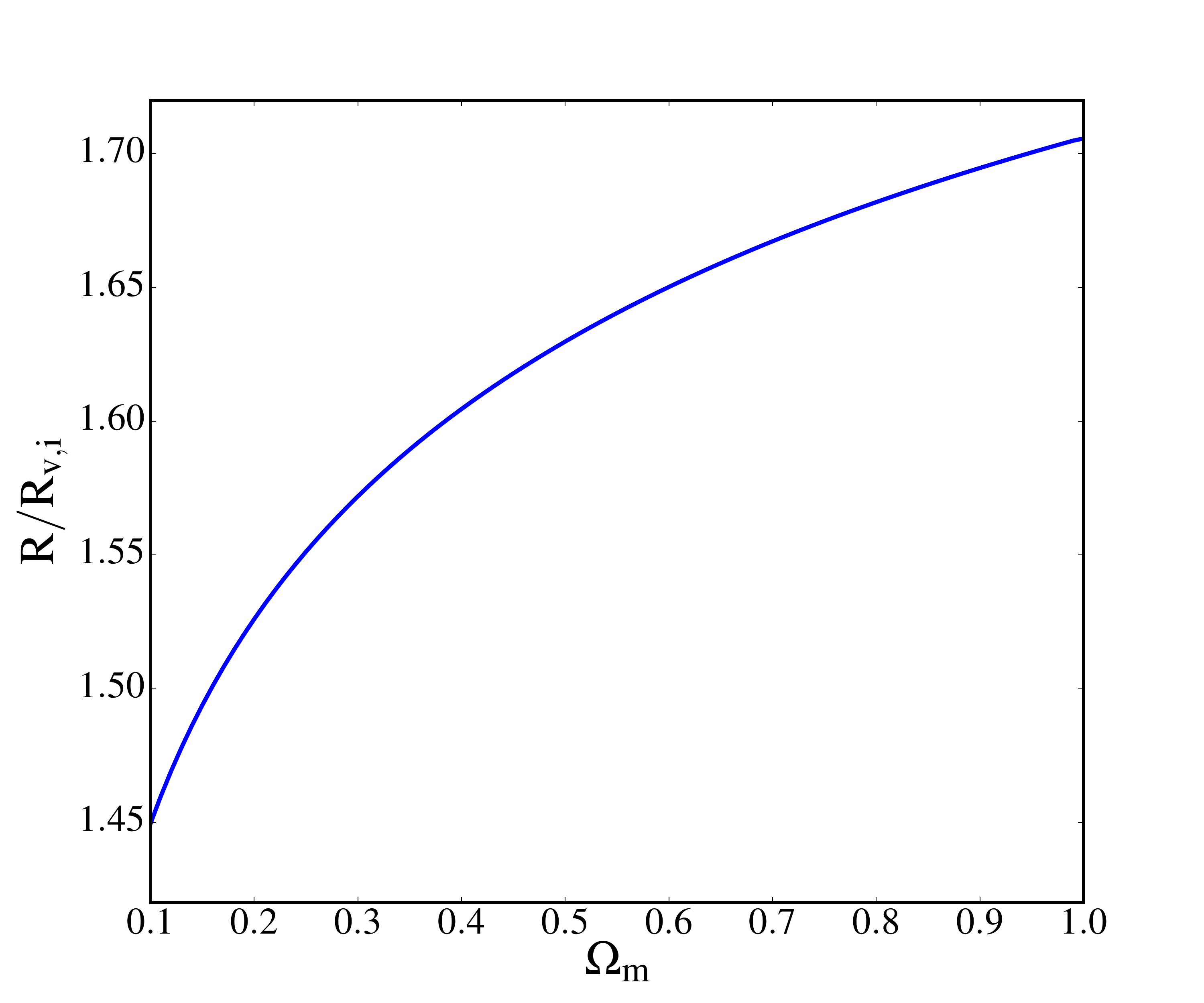}
	\caption{Comoving void radius at $a=1$ normalised by its initial size as a function of $\Omega_{\rm m}$ in flat $\Lambda$CDM universes.}
	\label{fig:Rpeak}
\end{figure}

In this section, we investigate the spherical evolution model in cosmologies with dark energy. We keep the inverse top-hat profile assumption for the initial density for the purpose of comparing solutions with those in an EdS cosmology.  
Switching from a $\Lambda$ = 0 to a $\Lambda$CDM cosmology, the dark energy term is added to the acceleration equation, yielding
\begin{equation}
\ddot{R} = -\frac{GM}{R^2} +\Omega_{\Lambda} H_0^2 R, 
\label{equation:lamacc}
\end{equation}
where $\Omega_{\Lambda}$ is the present day dimensionless density parameter for the cosmological constant $\Lambda$, and $\rm H_0$ is the present day Hubble constant. We chose the density parameters adopted in \cite{lietal2012} which are $\Omega_m = 0.24$ and $\Omega_{\Lambda} = 0.76$, for the purpose of comparing the model with voids in N-body simulations of the same cosmological parameters in Section \ref{sec:nbody}. The dark energy term is positive, counteracting the effect of gravity. The presence of dark energy acts as a damping term, suppressing the growth of the peculiar velocity compared to the case in EdS, as shown in the bottom panel of Fig~\ref{fig:LCDMtot}. This effect partly quenches the velocity gradient between the inner and outer shells, hence delaying shell-crossing. $\Lambda$CDM voids can therefore expand for longer without reaching the epoch of shell-crossing, as compared to EdS voids. This can be seen in the top panel of Fig~\ref{fig:LCDMtot} and in Fig~\ref{fig:deltavsa} where voids start from the same scale factor $a_i$ and initial density contrast $\delta_i$ in both the EdS and $\Lambda$CDM universes and are evolved to the same final redshifts. The two voids in different cosmologies follow closely to each other at the early times, but the evolution of the $\Lambda$CDM void slows at late times, having a relatively smaller void radius and smaller amplitude of density contrast at both the interior and the edge of the void. By $a=1$, the void in the EdS cosmology is about to reach shell-crossing. The comoving radius of the void in $\Lambda$CDM is smaller by $\approx 6$\%.  We compare the comoving void radii for different values of $\Omega_m$ at $a=1$ in flat $\Lambda$CDM universes in Fig \ref{fig:Rpeak}. Again, we find that the void radius decreases as the amplitude of the dark energy term increases.

\begin{figure}
	\centering
	\begin{subfigure}{0.5\textwidth}
		\includegraphics[width=1.0\linewidth]{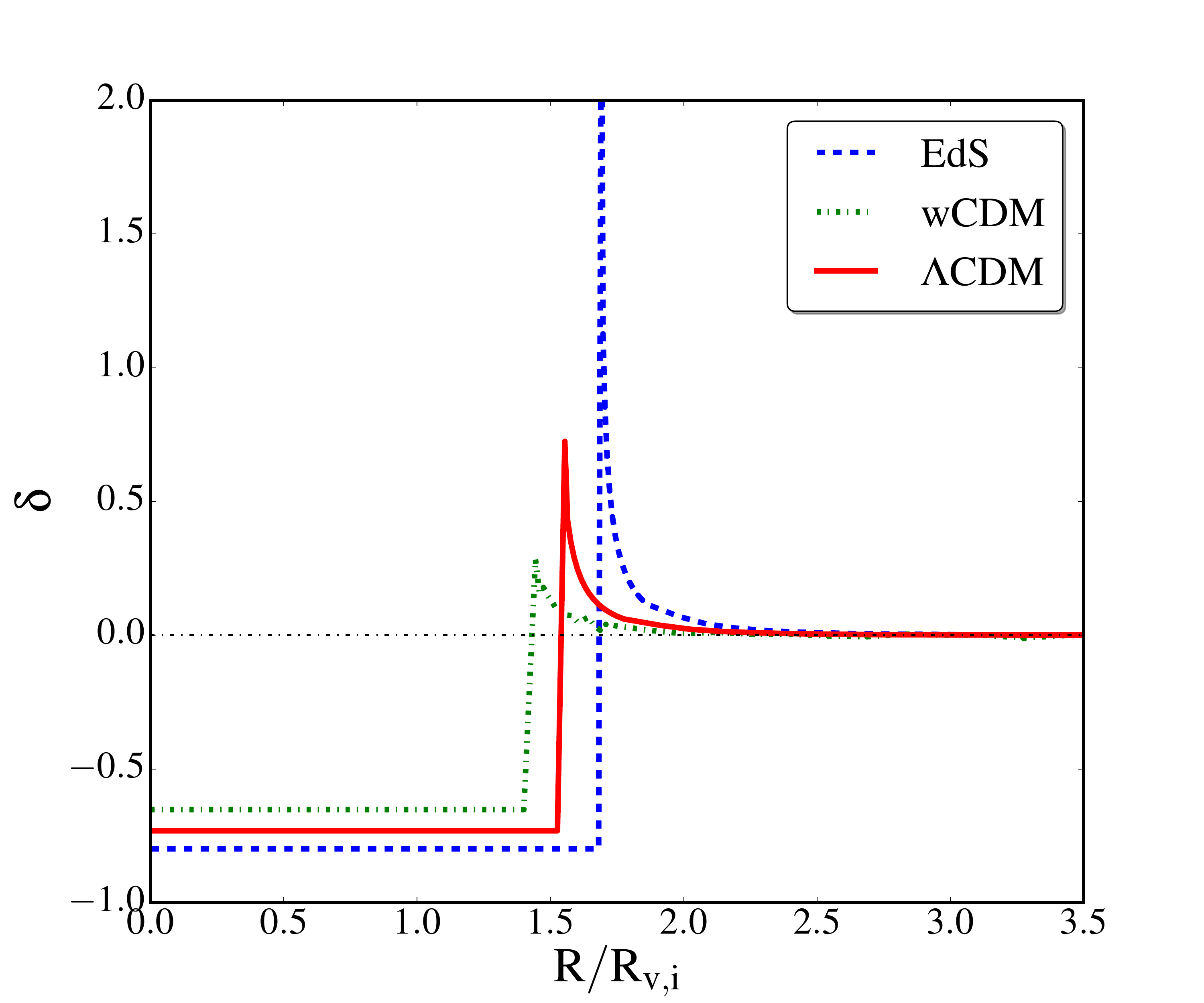}
	\end{subfigure} %
	\begin{subfigure}{0.5\textwidth}
		\includegraphics[width=1.0\linewidth]{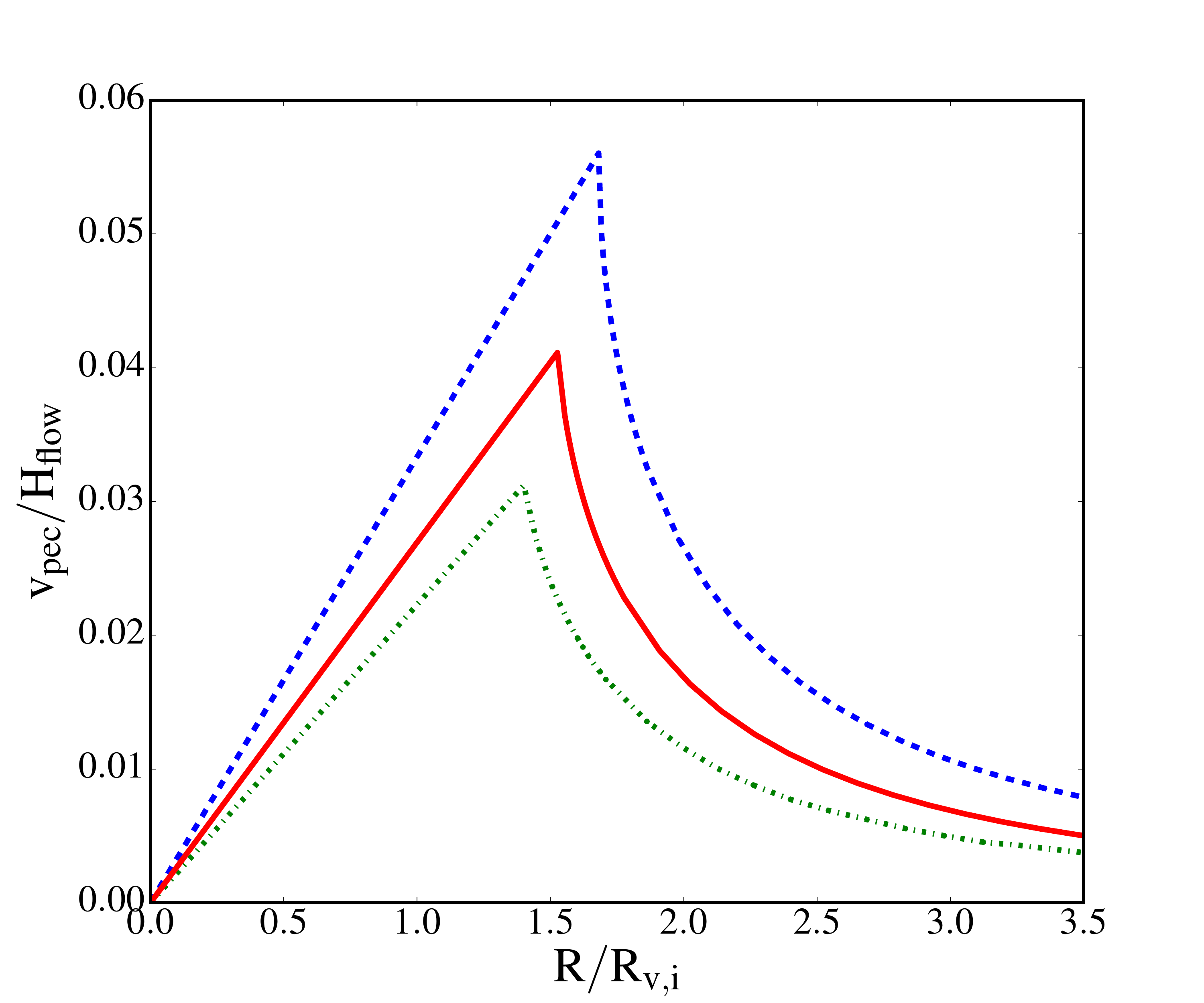}
	\end{subfigure} %
	\caption{\emph{Top:} The density contrast at $a = 1$ for three different cosmologies labelled by the legend. The fiducial $\Lambda$CDM model has $\Omega_{\Lambda}= 0.74$, and the $w$CDM model has $w=-0.5$, while the rest of parameters are the same as the fiducial model. \emph{Bottom:} Velocity profiles at $a=1$ for the same cosmologies normalised by the initial Hubble flow.}
	\label{fig:multicosmo}
\end{figure}

For general cases where the dark energy equation of state $w$ is not necessarily $-1$, Eq~\ref{equation:lamacc} becomes
\begin{equation}
\ddot{R} = -\frac{GM}{R^2} - \frac{\Omega_{\Lambda} H_0^2 R}{2}(1+3w)a^{-3(1+w)}.
\label{equation:qcdmacc}
\end{equation}
An example of void profile at $a=1$ for $w=-0.5$ is compared with the fiducial dark energy model, shown in Fig~\ref{fig:multicosmo}. With $w=-0.5$, the universe has been expanding faster than the case of $w=-1$ until $a=1$. The void experiences stronger background expansion from the dark energy term, which suppresses the development of peculiar velocities when compared to the fiducial $\Lambda$CDM case.  It therefore appears to be smaller and shallower at the interior. In contrast,  for $w<-1$, the void will be more evolved than the case in $\Lambda$CDM.  The distinction between different models of dark energy in terms of the density and velocity profiles suggests that voids have the potential to constrain dark energy parameters.

\begin{figure}
	\centering
	\includegraphics[width=1.0\linewidth]{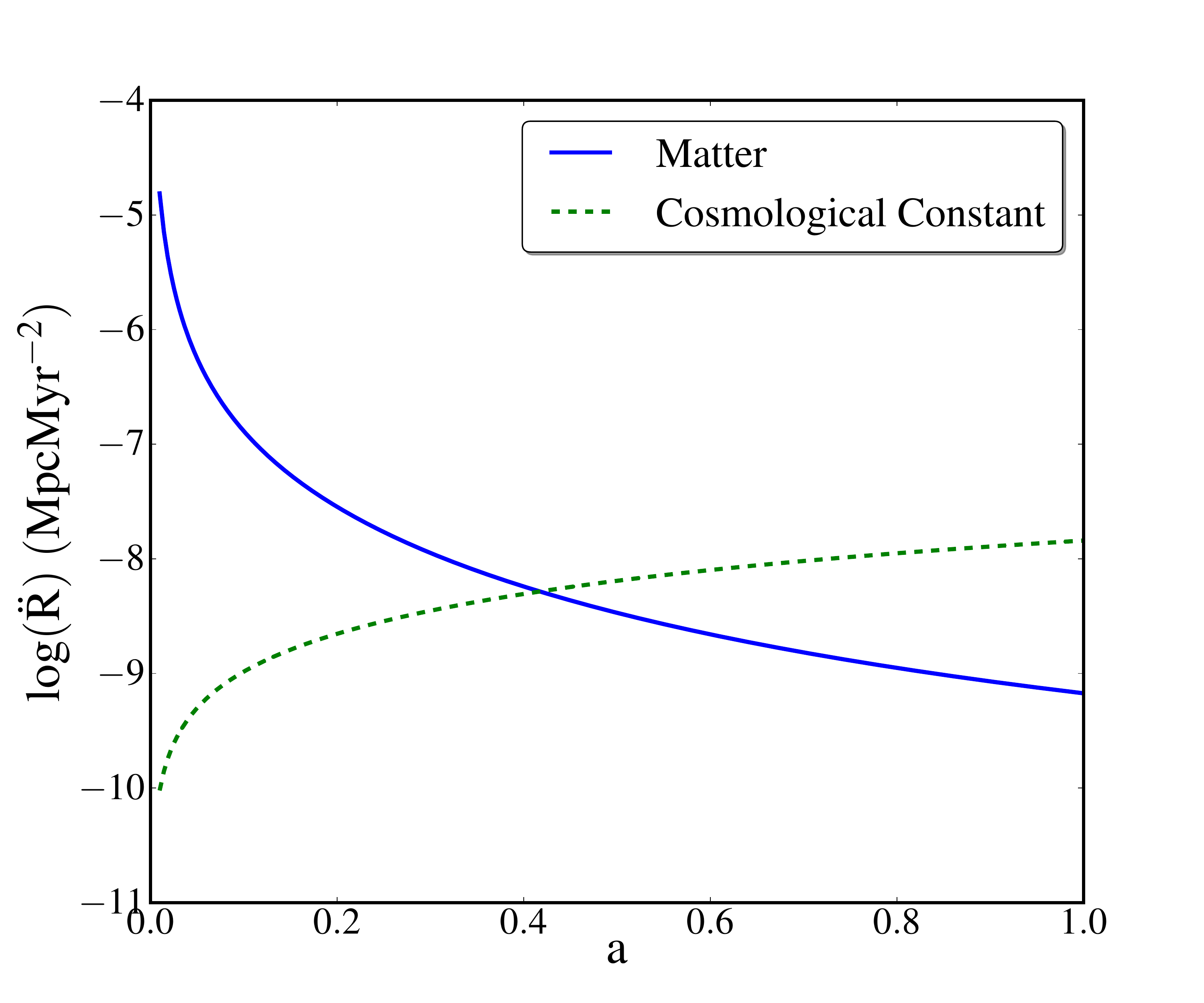}
	\caption{Contributions of dark matter (solid line) and dark energy (dashed line) to the acceleration of spherical shells shown in Eq \ref{equation:lamacc}, as a function of scale factor. The dark energy component dominates over the acceleration at $a \approx 0.4$, where the initial density contrast is chosen such that a void in EdS is on verge of shell crossing at $a=1$.  This value can be compared to the scale factor at which dark energy is dominant in our Universe, which is $a \approx 0.67$ \citep{friemanetal2008}.}
	\label{fig:LCDMforce}
\end{figure}

To further investigate the effect of dark energy on the expansion history of voids, we plot the contribution of acceleration from the mass part and dark energy part on the RHS of Eq~\ref{equation:lamacc}, shown in Fig~\ref{fig:LCDMforce}. It is interesting to see that the amplitudes of these two terms are equal at $a\approx0.4$ or $z\approx 1.5$ in voids. This is an earlier time than the epoch when dark energy starts to dominate the dynamics of the universe as a whole ($a \approx 0.67$ or $z \approx 0.5$). This is expected as the void region is a `bubble' with lower dark matter density as compared to the average of the universe. Since the dark energy density is thought to be the same regardless of matter density environment, it is more dominant in void regions and has been dominating for a longer time than in the universe as a whole. Because the dynamics of voids are affected more strongly and for a longer time by dark energy, they are a potentially powerful laboratory to test the nature of dark energy.

Finally, we have checked that with the same top-hat initial void profiles, when allowing the void to evolve to shell-crossing in the $\Lambda$CDM universe, 
the density contrast at shell-crossing is the same as that in the EdS. This occurs at a later epoch hence the proper physical radius of the void would be greater than its EdS counterpart.

\begin{figure}
	\centering
	\begin{subfigure}{0.5\textwidth}
		\includegraphics[width=1.0\linewidth]{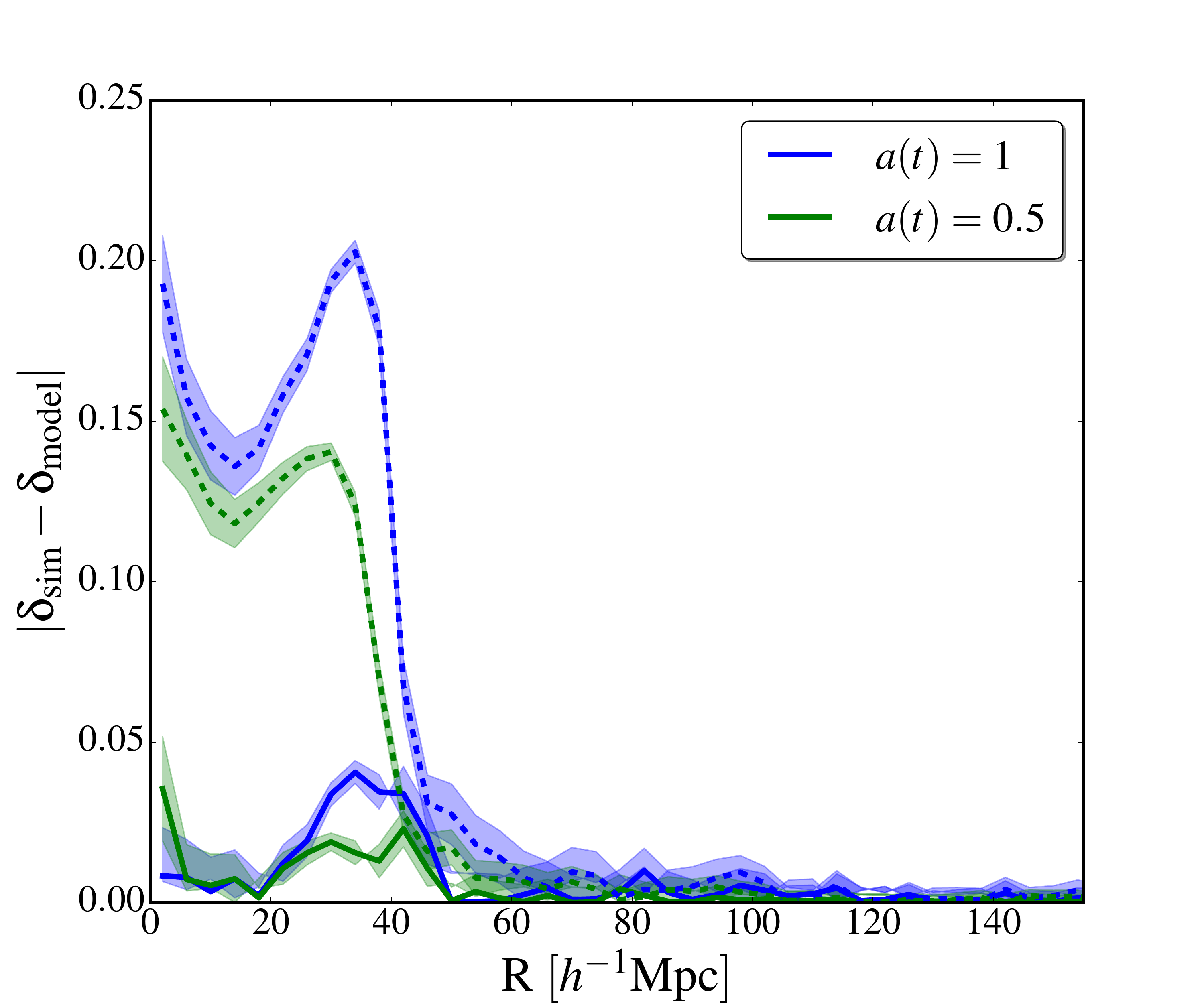}
	\end{subfigure} %
	\begin{subfigure}{0.5\textwidth}
		\includegraphics[width=1.0\linewidth]{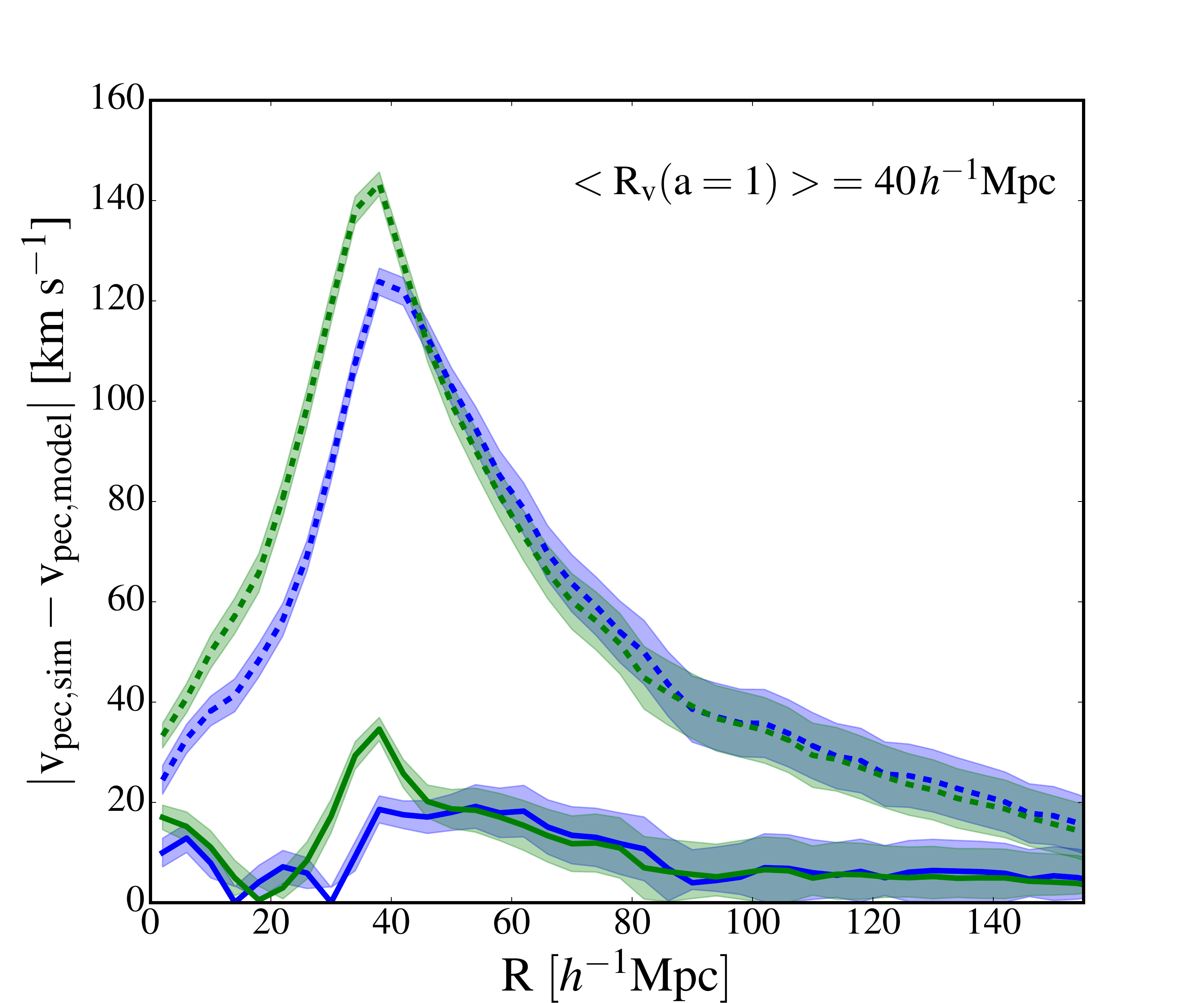}
	\end{subfigure} %
	\caption{Effect of various velocities as initial conditions on the absolute difference in density contrast (top) and peculiar velocity (bottom) profiles for a void with average an average radius of $40\mpcoh$.  The solid lines show the absolute difference if the initial velocity includes the peculiar velocity from the N-body simulation, whereas the dashed lines show the absolute difference if only Hubble flow is used as the initial velocity.  Blue and green represent $a=1$ and $a=0.5$, respectively.  The shaded regions represent a 1$\sigma$ errors on the N-body simulation curves.}
	\label{fig:velcomp}
\end{figure}

\begin{figure*}
\centering
\begin{subfigure}{0.475\textwidth}
\includegraphics[width=1.0\linewidth]{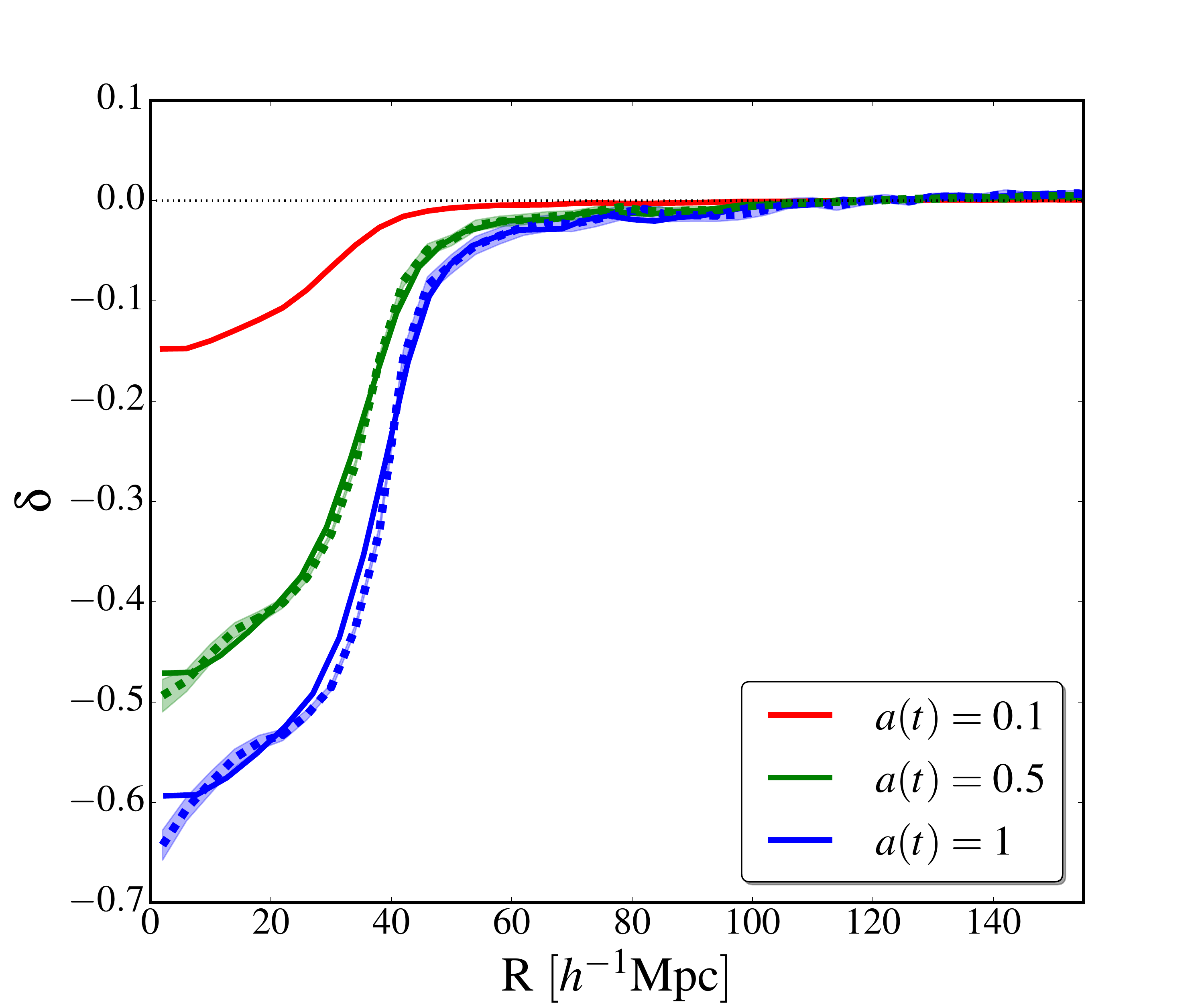}
\end{subfigure} %
\begin{subfigure}{0.475\textwidth}
\includegraphics[width=1.0\linewidth]{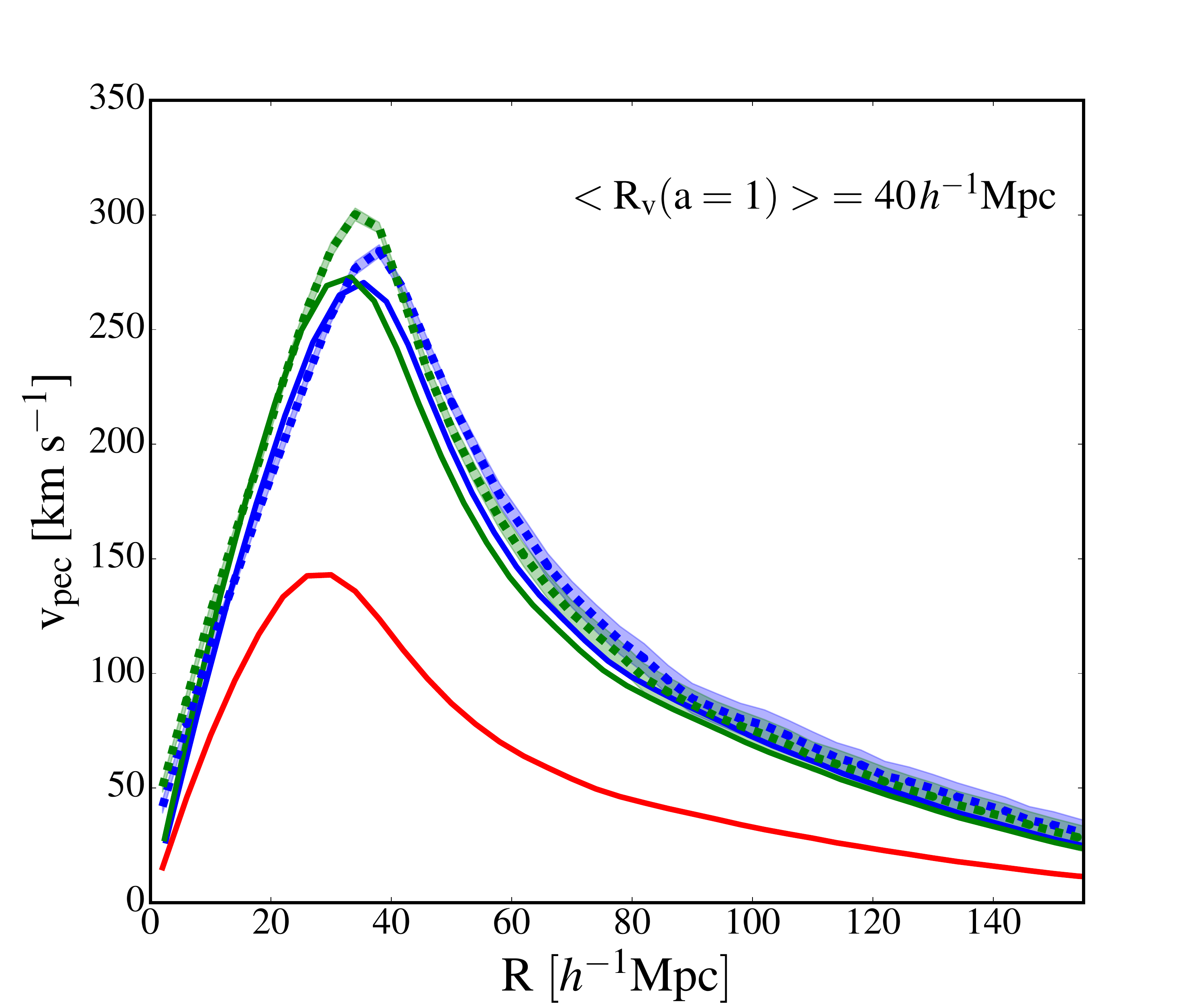}
\end{subfigure} %
\begin{subfigure}{0.475\textwidth}
\includegraphics[width=1.0\linewidth]{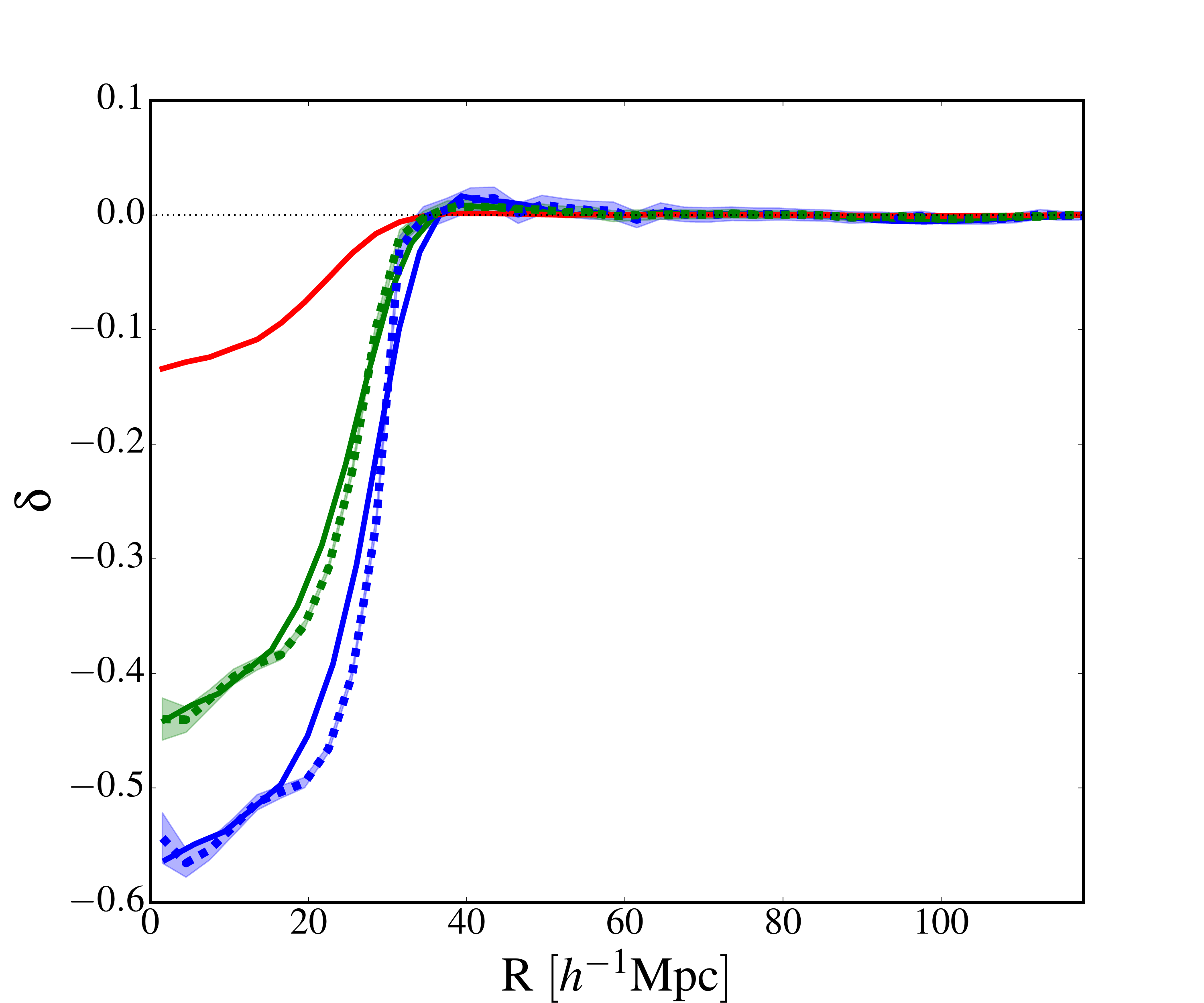}
\end{subfigure} %
\begin{subfigure}{0.475\textwidth}
\includegraphics[width=1.0\linewidth]{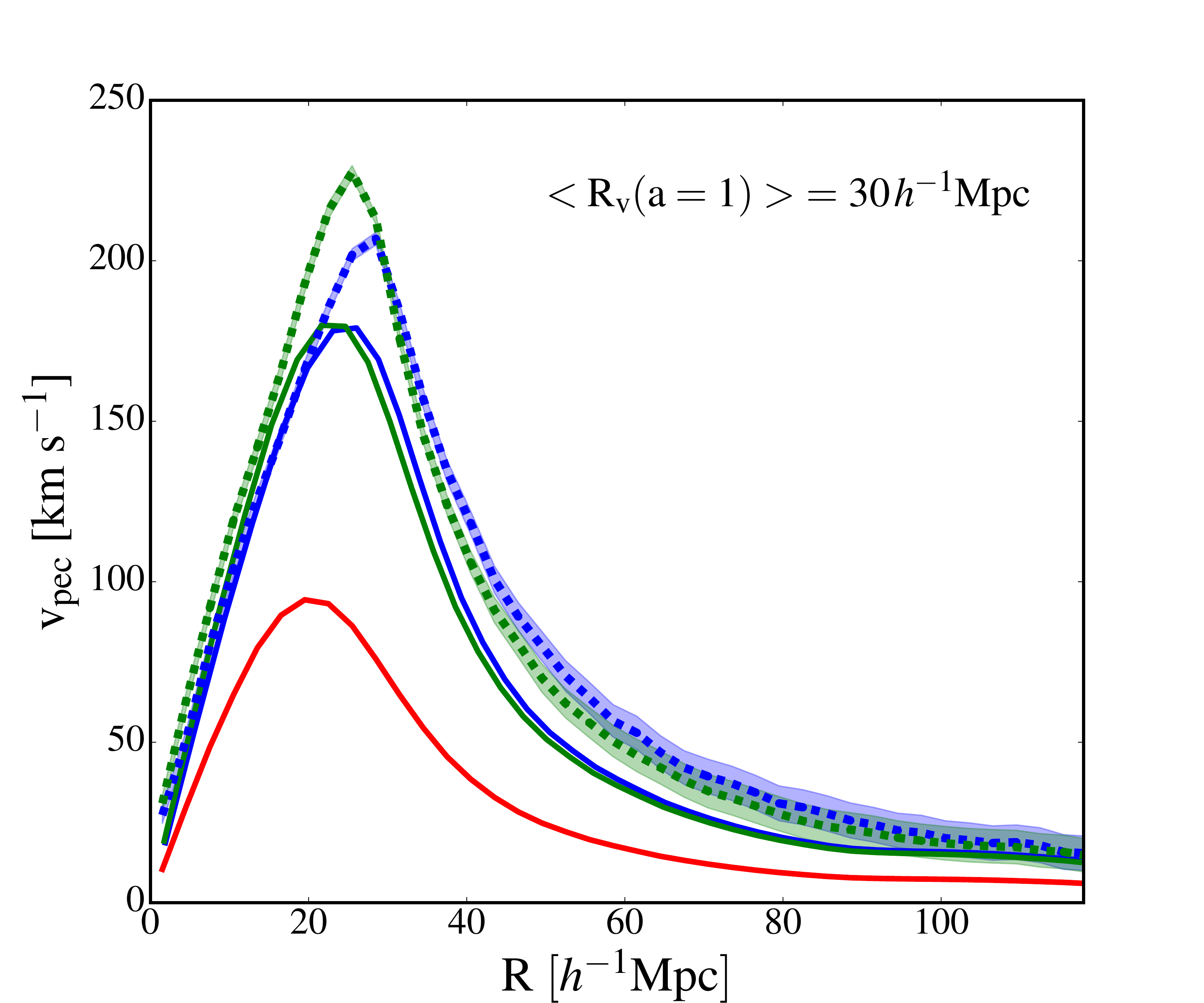}
\end{subfigure} %
\begin{subfigure}{0.475\textwidth}
\includegraphics[width=1.0\linewidth]{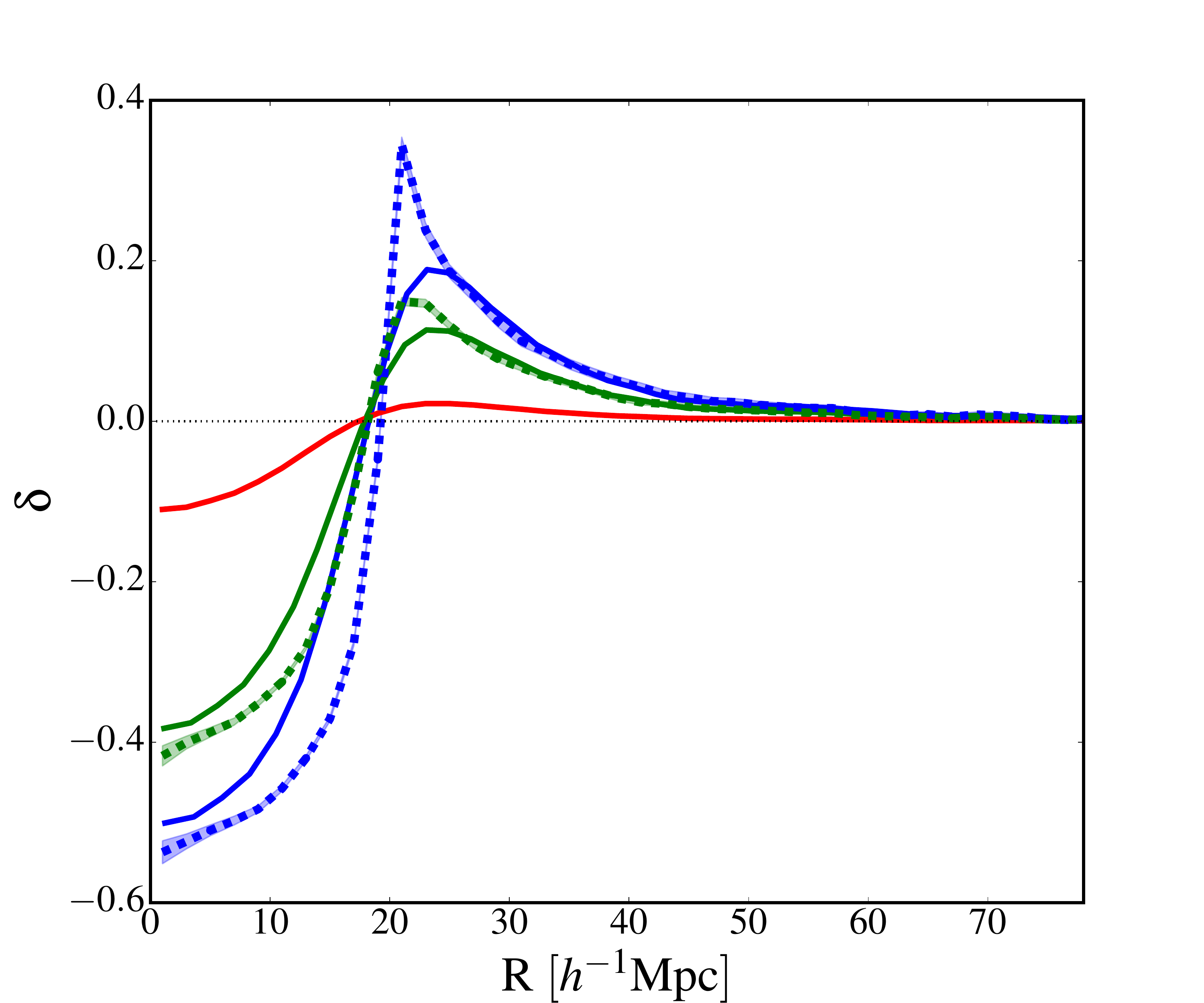}
\end{subfigure} %
\begin{subfigure}{0.475\textwidth}
\includegraphics[width=1.0\linewidth]{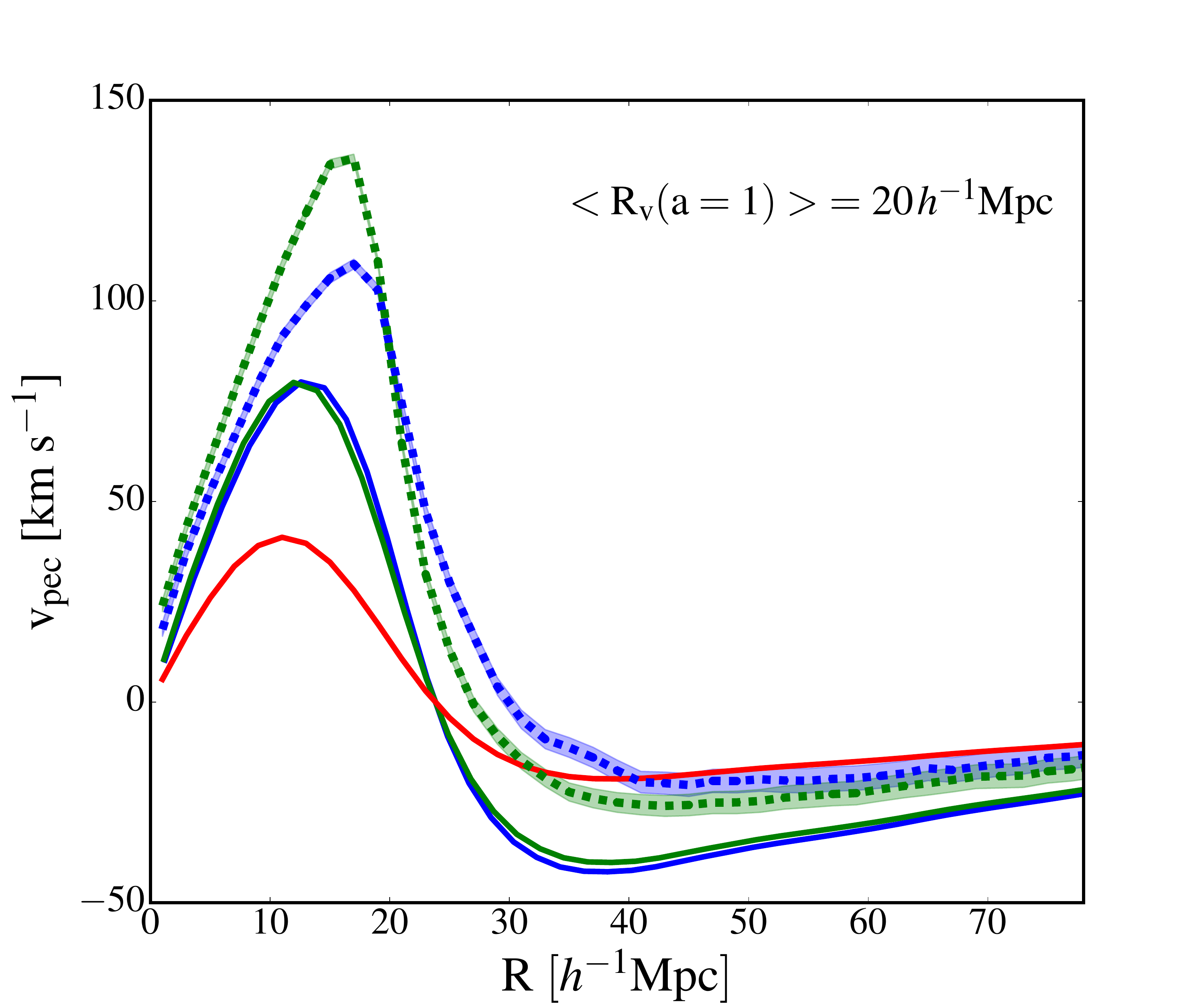}
\end{subfigure} %
\caption{Evolved void profiles in terms of density contrast (left) and peculiar velocity (right) predicted from the spherical model (solid) compared to  N-body simulations (dashed).  The rows from top to bottom represent voids with average radii of $40$, $30$, and $20\mpcoh$, respectively, at $a=1$. Solid curves of different colours represents results from model at different epochs labelled by the legend. The red curve at $a=0.1$ is the initial profile measured in the N-body simulations and used as input to the spherical model.  The shaded regions represent 1$\sigma$ errors on the N-body simulation curves.}
\label{fig:nbody}
\end{figure*}

\section{Comparison to N-body Simulation Results} \label{sec:nbody}
With the numerical solver for the spherical evolution model applied for different cosmologies in the previous sections, we now use it to solve the evolution 
for voids with initial conditions taken from N-body simulations. 

\subsection{N-body Simulation} \label{sec:nbodycomp}
We employ N-body simulations of a $\Lambda$CDM model with the
following parameters: $\Omega_{\rm m}=0.24$, $\Omega_{\rm
  \Lambda}=0.76$, $h = 0.73$, and $n_{\rm s}=0.958$ and
$\sigma_8=0.80$ from \cite{lietal2012}. The volume of the simulation box is
$(1$ ${h^{-1} \rm Gpc})^3$. We identify voids using all haloes above a 
minimum halo mass of $M_{\rm min}=10^{12.8}$ $h^{-1} \rm M_{\odot}$ to ensure that 
each halo contains at least 100 particles. Voids are found in the halo field with the
spherical underdensity algorithm described in \cite{caietal2015}, which
is based on the algorithm of \cite{padillaetal2005}. In the void algorithm, maximal spheres are grown
from a set of grid points, within which the number density of haloes
satisfies the criterion $\Delta \le 0.2$. Void candidates are ranked
in decreasing order of radius. Spheres that overlap with a neighbour
by more than 50\% of the sum of their radii are rejected. 
Technical details on the void catalogue can be found in \cite{caietal2015}. 

With the void centres defined at $a=1$ from simulations, we measure the dark matter density and velocity profiles around them.
To ensure that the voids from the simulations are close to spherical, we stack voids 
with radii in a narrow range of $40$, $30$, and $20\mpcoh$ at $a = 1$. We then use the same void centres in comoving coordinates to measure the stacked density and velocity 
profiles at $a=0.1$ and $a=0.5$. The density and velocity profiles at $a=0.1$ are treated as the initial conditions used by our numerical solver. 

Before proceeding to evolve the profiles, we have verified that the peculiar 
velocities measured from the simulation at $a=0.1$ can be accurately reproduced using the density profiles via the linear relation \citep{peebles1993}: 
\begin{equation}
v_{\rm pec}=-\frac{1}{3}aHf\bar\delta(r),
\label{equation:linvel}
\end{equation}
where $f \equiv d \ln D /d \ln a$ is the linear growth rate, $D$ is the linear growth factor, $H$ is the Hubble constant at $a$, and $\bar \delta(r)$ is the cumulative density profile from the model. 
The initial density and velocity profiles of our chosen voids  satisfy the linearised continuity equation and can be considered as linear. It is important to note that we need to include these non-zero peculiar velocities in our solver for the acceleration equation in order to obtain a sufficient level of accuracy in the density contrast profiles between the spherical model and N-body simulation.  Setting the initial peculiar velocity to zero for the analytical solutions with the top-hat model may seem reasonable since the peculiar velocity is usually negligibly small compared to the Hubble flow in the linear regime, however our results suggest that this is not the case.  This can be understood by the fact that N-body simulations use the total velocity as the initial condition \emph{i.e.} Hubble flow plus peculiar velocities, rather than just the Hubble flow alone.  Excluding peculiar velocities at the initial condition is equivalent to setting the initial growth rate of a void to be zero due to a cancellation of the growing and decaying modes \emph{i.e.}, $\dot{\delta}=0$.  The subsequent evolution of a void with this setting, assuming only the growing mode, will have a prefactor of 3/5 in the amplitude of density fluctuations compared to the case where the initial peculiar velocity is set according to linear theory.  Fig \ref{fig:velcomp} shows the $\Delta\delta$ (top) and $\Delta \rm v_{\rm pec}$ (bottom) between the N-body simulation and profiles which include an initial peculiar velocity (solid lines) and profiles that only use the Hubble flow (dashed lines) as the initial velocity.  As shown in Fig~\ref{fig:velcomp}, setting the peculiar velocity to be zero at the initial time largely slows down the evolution of the density and makes the predicted void profiles shallower than the simulation results.  In our analysis (Fig \ref{fig:nbody}) we use the linearly derived peculiar velocity (Eq \ref{equation:linvel}) plus the Hubble flow instead of the peculiar velocity from the simulation.  Although we find that using the peculiar velocity from the simulation as initial conditions for the model makes the results agree slightly better at small radii, we perform our analysis with the linearly calculated peculiar velocity because it is simple to obtain, requiring only a knowledge of the density contrast.

Having understood the effect of peculiar velocities in N-body simulations, we then calculate the evolved profiles at $a=0.5$ and $a=1$ and compare them to the simulation results. Fig~\ref{fig:nbody} depicts the density profiles and peculiar velocity profiles as a function of comoving void radius at the three epochs, where the dotted lines are the simulation and the solid lines are the model. We find that the void profiles at the initial time have a slower slope at their edges than that of a top-hat. For the relatively large voids, $\rm R_v=40$ and $30\mpcoh$, we find good agreement between the spherical model and the N-body simulations for the void density profiles at all epochs, as shown by the comparisons of the dashed curves versus the solid curves on the left-hand panel of Fig~\ref{fig:nbody}. For smaller voids with $\rm R_v=20$$\mpcoh$ however, the agreement between the spherical model and N-body simulations at the late time degrades. We suspect that this is due to mis-centering between the voids at the late time versus their initial conditions, caused by the void's non-zero bulk motions, \emph{i.e.} simulated voids may have been moving throughout their evolution history from $a=0.1$ to $a=1$; the amplitudes of such motions have been shown to be larger for smaller voids \citep{ceccarellietal2015}. This scenario would qualitatively explain the fact that the model prediction for smaller voids has a shallower density profile interior and a less-sharp density ridge, compared to that from N-body simulations at the late time. It may be possible to further improve the agreement between the model and simulation profiles at the late time for smaller voids, if one accurately tracks void centres back to their initial positions. We also suspect that smaller voids may be less spherical \citep{sheth} and more affected by tidal forces from their large-scale environments. This is not accounted for by the spherical evolution model hence it may diminish the agreement between the model and N-body simulations. We leave the investigation of small voids for future work.

Regarding peculiar velocities, the spherical model generally underpredicts their amplitudes by a few percent up to nearly 10\% at the peak of the outflow for $\rm R_v=40$ and $30\mpcoh$, and by a larger amount for $\rm R_v=20$$\mpcoh$. It might seem surprising that these deviations for the predicted peculiar velocity are not reflected as deviations in the predicted density profiles. This can be understood as follows: the evolution of the density profiles is determined by the total velocity (peculiar velocity plus Hubble flow).  At the scale of our interest, the Hubble flow dominates over the peculiar velocity, thus small deviations in peculiar velocities are inconsequential to the resulting density profile. Also, any differences will manifest themselves in the density profile integrated over a sufficiently long period of time, so we expect the difference in the density profile to show up at a later epoch as compared to when the difference starts to emerge in the velocity profiles.

It is worth noting that if one simply applies linear theory to evolve the density profiles from the initial conditions, the amplitudes of the density profiles are largely overpredicted. This suggests that the spherical model successfully describes the dynamics for large voids, \emph{i.e.} the growth of the density contrast in voids has to slow down and is significantly slower than predicted by linear theory. We also note that even though the initial peculiar velocities seem negligibly small compared to the Hubble flow, they need to be included in our solver for the acceleration in order to obtain a sufficient level of accuracy in the density contrast profiles between the spherical model and N-body simulation.

\section{Conclusions} \label{sec:conclusion}

We investigate the spherical evolution model for voids in different cosmologies and compare voids in Einstein-de Sitter, $\Lambda$CDM, and $w$CDM cosmologies. We start with the assumption that the initial density of voids can be modelled with an inverse spherical top-hat profile. We find that the presence of dark energy damps the effect of gravity sourced by dark matter and suppresses the growth of peculiar velocities. This causes the same void to decrease in size by a few percent when comparing EdS to $\Lambda$CDM at the epoch when shell-crossing is about to occur in the EdS universe. In general, the impact of dark energy for the evolution of voids increases as the dark energy density increases relative to the dark matter component. This implies that its impact is stronger for voids than for the whole universe on average. The dynamics of voids have been affected by dark energy for a longer time and therefore the imprint of dark energy is stronger within them. This makes voids potentially powerful candidates for constraining dark energy.

With the success of the generalised model demonstrated, we compare the model to N-body simulations. Using the initial conditions from the simulation, we evolve voids of different sizes using the spherical model and compare the final density and peculiar velocity profiles with simulation outputs. We show that the model successfully reproduces the density and velocity profiles for voids with radii of $30$ and $40\mpcoh$, with the agreement for the velocity profiles being slightly worse, \emph{i.e.} the model underpredicts results from simulations by a few percent and up to 10\% at the peak.  The success of the spherical model for tracking the evolution of large voids opens up the possibility of using it to constrain dark energy. The performance of the model is not as successful for smaller voids, which may be due to mis-centring errors in our determination of the position of void centres within the simulation when the initial conditions are measured with the evolved late time void centres.

\section*{Acknowledgements}

 VD acknowledges the Higgs Centre Nimmo Scholarship and the Edinburgh Global Research Scholarship.  YC was supported during this work by funding from an STFC Consolidated Grant. CH acknowledges support from the European Research Council under the grant number 647112.  YC and JAP were supported by the European Research Council under grant number 670193.  We thank Baojiu Li for providing the N-body simulation for this analysis.
\\
\\




\bibliographystyle{mn2e}
\bibliography{\myref}
\appendix

\bsp	
\label{lastpage}
\end{document}